\documentclass[conference]{IEEEtran}
\usepackage[utf8]{inputenc}
\usepackage[frozencache,cachedir=.]{minted}

\IEEEtriggeratref{21}

\usepackage{cite}
\usepackage{amsmath,amssymb,amsfonts}
\usepackage{algorithmic}
\usepackage{graphicx}
\usepackage{textcomp}
\usepackage{xcolor}
\usepackage{url}
\usepackage[normalem]{ulem}
\begin{document}

\title{QuickREST: Property-based Test Generation of OpenAPI-Described RESTful APIs}

\author{\IEEEauthorblockN{Stefan Karlsson}
\IEEEauthorblockA{ABB AB, Mälardalen University\\
Västerås, Sweden\\
Email: stefan.l.karlsson@\{se.abb.com, mdh.se\}}
\and
\IEEEauthorblockN{Adnan \v{C}au\v{s}evi\'{c}}
\IEEEauthorblockA{Mälardalen University\\
Västerås, Sweden\\
Email: adnan.causevic@mdh.se}
\and
\IEEEauthorblockN{Daniel Sundmark}
\IEEEauthorblockA{Mälardalen University\\
Västerås, Sweden\\
Email: daniel.sundmark@mdh.se}}

\maketitle

\begin{abstract}
RESTful APIs are an increasingly common way to expose software systems functionality and it is therefore of high interest to find methods to automatically test and verify such APIs. To lower the barrier for industry adoption, such methods needs to be straightforward to use with a low effort.
This paper introduces a method to explore the behaviour of a RESTful API.
This is done by using automatic property-based tests produced from OpenAPI documents that describe the REST API under test. We describe how this method creates artifacts that can be leveraged both as property-based test generators and as a source of validation for results (i.e., as test oracles). Experimental results, on both industrial and open source services, indicate how this approach is a low effort way of finding real faults. Furthermore, it supports building additional knowledge about the system under test by automatically exposing misalignment of specification and implementation. Since the tests are generated from the OpenAPI document this method automatically evolves test cases as the REST API evolves.

\end{abstract}


\section{Introduction}

Representational state transfer (REST) is an architecture style, introduced by Fielding, that describes constraints on web services \cite{fielding2000}. With REST, a system resource is exposed by a URI and created, read, updated and deleted with HTTP verbs. A service that uses the REST architecture is said to be \textit{RESTful}. REST APIs are a common way of exposing web services on the internet. The web-site \textit{Programmableweb\footnote{https://www.programmableweb.com}} contains more than 20,000 APIs in their directory\footnote{https://www.programmableweb.com/apis/directory}. This includes well known services such as Twitter\footnote{https://twitter.com/}, YouTube\footnote{https://www.youtube.com/}, Facebook\footnote{https://www.facebook.com} and cloud providers such as Microsoft Azure\footnote{https://azure.microsoft.com} and Amazon Web Services\footnote{https://aws.amazon.com}.
REST APIs are also commonly used when exposing internal interfaces in a microservice architecture \cite{Dragoni2017}. Test methods targeting REST APIs can thus be useful both on the internal and external interfaces of a software system.

The automatic exploration of RESTful APIs has the potential to save developer and tester effort, get insights from the system under test (SUT) and fill the gaps of lack of tests where such gaps exist. Recent results by Atlidakis et al. \cite{Atlidakis:2019:RSR:3339505.3339600} and Arcuri \cite{Arcuri:2019:RAA:3292526.3293455} have shown that fault finding can be done automatically for REST APIs, both as a black- or white-box approach. We think such approaches could help developers and testers in exploring functionalities of REST APIs, particularly if more properties are evaluated other than HTTP status codes.

In our experience, exploring functionalities of a SUT is mostly a manual effort in industry, and for good reasons. Some insights about a SUT need a human in the loop, such as evaluating if a usability pattern makes sense in the given domain. However, it may be valuable to automate the parts of exploration that can be done by machines, and in doing so freeing up human effort to be spent where it is of best use. 

REST APIs are increasingly commonly described with OpenAPI \cite{openapi}, which aims to standardize how RESTful APIs are described. Several frameworks for building REST APIs also include OpenAPI support. The OpenAPI document specifying REST APIs opens up an interesting way for an automatic tool to interact with the SUT, thus serving as a possible interaction model for automatic exploration of REST APIs. 

However, the interaction model is only one piece of the puzzle. Methods for generating input data to tests and some way to evaluate the results, an oracle, are also needed. In addition, to have a useful technique in industry, a meaningful and efficient test reporting is also of high importance. If analyzing the results from automated testing is associated with a high cost, potential adoption in industry might be rather low.

In this paper, we propose a method to leverage OpenAPI documents to automatically generate tests. Test inputs are generated using a two-fold mechanism: (i) randomly generated values that are agnostic to the specification, as well as (ii) randomly generated values that conform to the parameter specification in the given OpenAPI document. Test oracles, used to provide verdict on the conformance of response data, are also automatically generated from the OpenAPI document. They assert the REST API results in a property-based fashion. 

The choice of property-based testing is made due to its suitability for random exploration, the availability of libraries in that domain, and the feature of shrinking. Shrinking is a feature that, when a test case fails, aims at producing the smallest failing example possible. Property-based testing also allows for formulating and checking several properties of the test results such as if the response body conforms to a given specification, resulting in a stronger oracle than only asserting on the status code of an HTTP response.

In order for our method to be suitable to a wide range of REST APIs, with as little developer effort as possible, we develop the method as a black-box approach. The approach described in this paper has been evaluated in an industrial case study at ABB and on the large open-source software (OSS) GitLab\footnote{https://about.gitlab.com/}. 
Experimental results show that we can find faults and gain insights of the SUT given an OpenAPI document in an automated way.

To summarize we make the following contributions:
\begin{itemize}
    \item We introduce a method to help developers and testers automatically explore RESTful APIs with low effort.
    \item We describe how the combination of available open source libraries can be used to implement such a method, as well as provide a proof-of-concept implementation, thus lowering the barrier of entry for industry.
    \item We investigate how the configuration of generators effect the status code coverage and fault finding probability. 
    \item We present the results of applying the method to multiple APIs developed and used in industry as well as in OSS.
\end{itemize}

\section{Background}
Before going into the details of this work, we would like to provide some context and explanations of the terms used. 

\subsection{OpenAPI}
OpenAPI, also known as Swagger, is a way of describing REST APIs \cite{openapi}. OpenAPI \textit{specification} describes the format which needs to be followed by an OpenAPI \textit{document}. OpenAPI \textit{document} describes an instance of a REST API whose description follows the OpenAPI \textit{specification}.

Since much of the documentation and tooling around OpenAPI refer to Swagger it is worth to mention the history of the names. Swagger was the original name prior to OpenAPI, when the ownership was given to the OpenAPI Initiative \cite{openapi-history}. The Swagger name is still used to refer to the tooling as well as the OpenAPI specification. However, in this paper, we further refer to it as OpenAPI. We have also chosen to target the 2.0 version of the OpenAPI specification \cite{openapi-spec-2}. The reason to not use the later, 3.0 version, is that the 2.0 is still largely in use and it is the version used in the industrial system which is part of our evaluation. 

An OpenAPI-described REST API contains a set of available HTTP verbs (GET, POST, PUT etc.), the parameters and the responses. The parameters describe where the parameter is used in the request (path, query, body etc.), its name, type, and format. Responses are indexed by an HTTP return code and provide details of the structure of the response, if it is a value, an array or a complex object.

\begin{figure}[h]
    \centering
\begin{minted}{json}
{"swagger": "2.0",
 "info": {
   "version": "1.0.0",
   "title": "Example Service API",
 "basePath": "/theservice",
 "paths": {
   "/api/v1/objects/{objectId}": {
     "get": {
       "tags": [],
       "summary": "Object model lookup",
       "parameters": [
         { "name": "objectId",
           "in": "path",
           "required": true,
           "type": "string",
           "format": "uuid"}],
       "responses": {
         "200": {
          "schema": {
            "$ref": "#/definitions/ObjectInfo"}}}}}},
 "definitions": {
   "ObjectInfo": {
     "type": "object",
     "properties": {
       "objectId": {
         "format": "uuid",
         "type": "string"},
       "name": {
         "type": "string"}}}}}}
\end{minted}
    \caption{OpenAPI document in JSON-format}
    \label{fig:open-api-doc}
\end{figure}

Figure \ref{fig:open-api-doc} shows a simple example of an OpenAPI \textit{document} described in JSON format, which is following the 2.0 version of the OpenAPI \textit{specification}. We can see examples of the available paths and its parameters and expected results. In addition, there is a small definition that specifies the structure of a data type, called \textit{object} in this example.  

\subsection{Property-based testing} \label{property-based testing}
The main idea of property-based testing (PBT) is to generate input data and to check if defined properties hold when exercising the SUT with that input. To get a better understanding of the basic principle, let us look at an example of using PBT to test a \textit{sort} function, which sorts a given list of numbers. The first step would be to generate a random list of input numbers. Such a generator is most likely included in a PBT library, but if not, the means to create custom generators are typically there. When we know that we can generate input data we can formulate a property of the function. An invariant of the sort function is that the number at position \textit{n} in the resulting list should be less than or equal to the number at position \textit{n+1}. To make this into a test we will \textit{check} the property. Checking will exercise the SUT for a given number of iterations. Each iteration is given input data from our generator and verifies that our defined property holds for each such input.

The feature of \textit{shrinking} is also common in PBT. This means that whenever random data has been generated to which a property has failed, the library will try to find the smallest input that fails the property in the same way by \textit{shrinking} it. An example of shrinking is provided later on in Section~\ref{sec_shrinking}.

Formulating properties can be challenging. One such challenge is how to make a model of the SUT without re-implementing the functionality of the property to be tested.
This challenge can be avoided by instead of trying to model the implementation, formulating invariants or symmetries.
Examples of invariants of a sorting algorithm are that the input and the output should be of equal length and that all input values should exist in the output. These invariants are true given any randomly generated input. Further, depending on the problem at hand we might be able to find symmetries that should hold. A symmetry is when the input data is invariant given a set of operations. As an example, consider a lossless compression algorithm. 
If we start with generated data and then compress and decompress the data, the symmetry of the compression should ensure that we end up with the same data that we started with.

QuickCheck~\cite{Claessen:2011:QLT:1988042.1988046} is known as the first tool for property-based testing. Today there exist re-implementations of QuickCheck in a number of languages, reaching multiple platforms. A few such examples are PropEr \cite{Papadakis:2011:PIT:2034654.2034663} for Erlang, ScalaCheck\footnote{https://www.scalacheck.org/} for Scala on the JVM, and FSCheck\footnote{https://fscheck.github.io/FsCheck/} for F\# on .NET. Our implementation uses the Clojure variant called TestCheck\footnote{https://github.com/clojure/test.check}. TestCheck contains basic generators and also the components needed to define own more complex generators. A reason to choose TestCheck is that Clojure.spec can produce TestCheck generators out of data specifications.

\subsection{Clojure.spec}
Clojure.spec\footnote{https://clojure.org/about/spec} is a library for the Clojure\footnote{https://clojure.org/} functional programming language. Clojure.spec provides functionality to define specifications of data and to validate if given data conforms to such a specification. These specifications are referred to as \textit{specs}. Clojure.spec also provides means to produce random data generators given a spec. As we shall see, this gives us leverage in our approach when we can use specs to both generate random input data and to validate responses.

As a small example to give an intuition of the basics of a spec, here is an example of a spec called ::age that defines that to be a valid age the given data must be a natural integer\footnote{Inclusive of 0, according to ISO 80000-2} and less than 150.
\mint{clj}|(s/def ::age (s/and nat-int? #(< % 150)))|

The simplest form of validation is to use the \texttt{valid?} function. The result will be a boolean representing if the data is valid checked against the spec.
\mint{clj}|(s/valid? ::age 15) => true|

To use this spec as a generator we first create a generator from the spec with \texttt{gen}. This returns a generator for that spec that can be used in any place where a generator is needed. Here we use it to produce 10 sample values.
\mint[fontsize=\footnotesize]{clj}|(sample (s/gen ::age)) => (0 0 2 1 1 10 16 1 26 8)|

The benefits of using Clojure.spec is both that we can use it as a random data generator and that we can leverage the same specs for validation.

\section{Proposed method}

We propose a method that for a given specification, OpenAPI in our case, produces input generators that are used in property-based tests as well as produce automatic oracles. The properties checked in the tests, with previously produced input generators, are a combination of predefined static properties and properties automatically derived as valid responses from the specification. Static properties are provided as status codes since success/failure codes are built into HTTP, while valid responses are specified in the OpenAPI document. This produces an automatic oracle that is able to validate if response from the SUT conforms to what is specified, as well as able to identify discrepancy in the given OpenAPI document and response codes received from the SUT.

We aim for a black-box method for test generation of REST APIs, that is as automated as possible. The reason for choosing an automatic black-box approach is to make it easy for developers and testers to use, and for the method to apply to any system using an OpenAPI documented API. We are thus not constrained by implementation details such as languages and platforms. 

The proposed approach also makes it straightforward to guide input when it is needed, by having custom generators. This allows testers and developers to guide the input in a way that makes sense for a given SUT. For example, the distribution of string generation between any kind of strings and pattern-based strings, like alpha-numeric strings or UUID (Universally Unique IDentifier), can be changed. By doing so, we can ensure that the SUT receives both valid and invalid inputs. A high frequency of any kind of strings would test the SUT in a more fuzzy approach while generating valid input expose a higher number of functionalities in the SUT to be tested.

In addition, our method proposes the \textit{bidirectional} use of the specification. By doing so, a specification is used both to generate valid input as well as to generate a verdict in the form of an automated oracle. The oracle is used to validate that responses conform to the given specification. Since this method automatically derives tests and oracles from the specification of the API under test, both the tests and oracles will automatically evolve with any changes to the API and its specification.

To make results approachable and useful, we propose to leverage the already available method of shrinking, where any property that fails will produce a result that is shrunken to its smallest reproducible case.

Furthermore, we seek to support additional test objectives to fault finding during the testing of the API. Finding faults and failures while fulfilling coverage criteria are considered traditional goals of a testing process and we would like to extend this goal with our method. The properties of an API that are validated when running our generated tests include not only correctness measures but can also check, for example, that all received responses are part of the specified behaviour.
Such properties give the ability to draw insights of how the SUT behaves. If a response code is received that is not part of the published OpenAPI document, then the behaviour of the SUT and the specification have diverged. While this may not be a fault in the system's behaviour, it is nevertheless an important insight. 

In essence, running tests that only result in a \textit{pass} or \textit{fail} result is a missed opportunity to gain new knowledge. If a test passes 100\% of the time, over time the probability of a failure will be lowered for each test execution, a point will be reached where no new knowledge is gained about the SUT with the computing effort given. As stated by Feldt, such a test case, still active but not effective for finding failures, has grown old \cite{6823896}. We try to solve this lack of knowledge gathering by generating new test cases that evolve with the selected interaction model, when the OpenAPI document describing the service evolves so does the tests generated, and to not only focus on fault finding properties. 

\section{Implementation}
To evaluate our proposed method, we implemented a proof-of-concept tool, called QuickREST \cite{replication}. QuickREST leverages our method and applies it to REST APIs that are specified with an OpenAPI document.

The overarching process of QuickREST when applying our method to REST APIs, which will be described in detail in following subsections, is as follows:
\begin{enumerate}
    \item Acquire the OpenAPI document, via HTTP or a file.
    \item Parse the JSON document to an internal format. \label{item:parsing}
    \begin{enumerate}
        \item Attach specifications for the parameters.
        \item Attach specifications for the responses.    
    \end{enumerate}
    \item Generate specifications for data definitions. \label{item:gen}
    \item Make test generators based on the specifications.
    \item Check the properties for each of the HTTP verbs defined.
    \begin{enumerate}
        \item If in a stateful sequence of operations, store the response.
        \item Select next input from the collected responses.    
    \end{enumerate}
    \item Report the test result.
\end{enumerate}

\begin{figure}
    \centering
    \includegraphics[width=\linewidth]{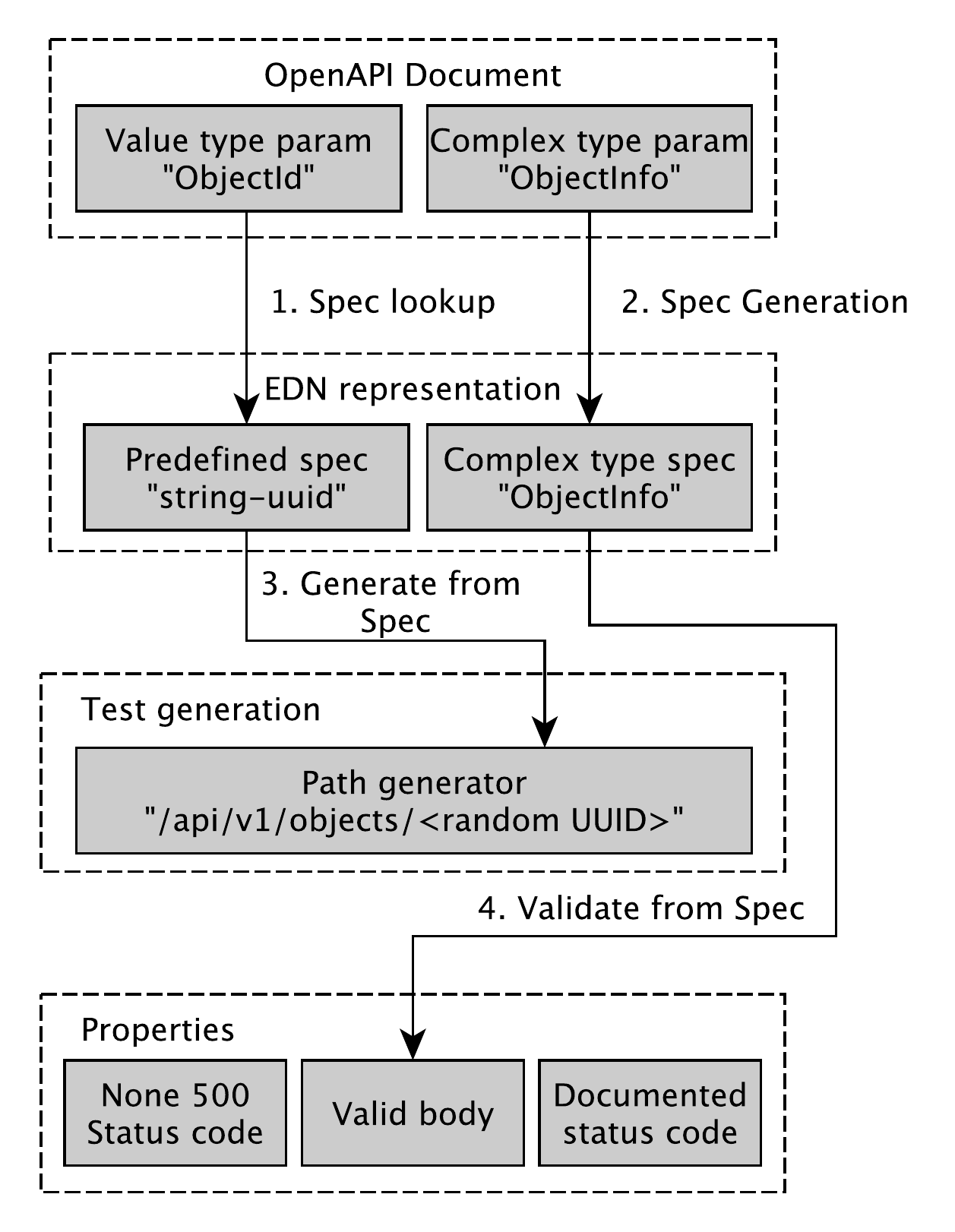}
    \caption{Example of creation and usage of Specs}
    \label{fig:spec-usage}
\end{figure}

Figure \ref{fig:spec-usage} is an example of how a REST API, as described in Figure \ref{fig:open-api-doc}, would be processed and tested.

\subsection{Parse the JSON document}
The first step of parsing is to represent the JSON data of the OpenAPI document in a format that is more suitable to work with in the language of choice. The JSON data is translated into extensible data notation\footnote{https://github.com/edn-format/edn} (EDN), which is a subset of Clojure, making the data of the OpenAPI document very accessible in a Clojure program.

The second step of parsing is to attach \textit{Clojure.spec} specifications to the defined parameters and responses. The attached specs will be used both for input generation and for result validation. Figure \ref{fig:spec-usage} shows an example of this where "ObjectId" (12 - 16 in Figure \ref{fig:open-api-doc}) and "ObjectInfo" (22 - 29 in Figure \ref{fig:open-api-doc}) are attached and used.

The OpenAPI \textit{specification} for parameters and responses state that the type can be a basic value-type such as number, boolean etc. or a schema reference to a complex object.

For basic value types, predefined specifications were made in the implementation. During parsing, the parameter/response type is then mapped to the spec for that specific type and format. For example, the OpenAPI parameter type of \texttt{string} with the type of \texttt{UUID} will map to the predefined spec shown in Figure \ref{fig:string-uuid}.
This \textit{spec} specifies (line 4) that the input should be a string and match the regular expression of a UUID. 
The \texttt{::string-uuid} spec has a custom generator that generates a \texttt{UUID} and then makes a \texttt{string} from that \texttt{UUID}. We need a custom generator in this case since the default one would be a string generator. The likelihood that the default generator would generate a random string that actually matches the UUID regex is very low and generation would fail.

The types in the OpenAPI document that are references to complex objects will be parsed and a reference to a spec is attached. The definition in Figure \ref{fig:complex-object} that references an \texttt{array} of \texttt{ObjectInfo} will be attached as a reference to the spec \texttt{:array/ObjectInfo}. In the case where the reference is to a single object and not to an array, as the 200 response in Figure \ref{fig:open-api-doc}, the result would be \texttt{:definitions/ObjectInfo}.

In summary, the parsing step of our process starts with the OpenAPI document in JSON format and results in the OpenAPI document represented in a format suitable for further processing, specifications attached to basic value types and reference types, for both parameters and responses.

\begin{figure}
    \centering
\begin{minted}{json}
{"200":
  {"description":"The query request was successful.",
   "schema":
     {"uniqueItems":false,
      "type":"array",
      "items":{"$ref":"#/definitions/ObjectInfo"}}}}
\end{minted}
    \caption{Definition of complex object reference}
    \label{fig:complex-object}
\end{figure}

\begin{figure}
    \centering
\begin{minted}{clj}
(s/def 
  ::string-uuid 
  (s/with-gen
    (s/and string? #(re-matches uuid-regex %))
    #(gen/fmap str (s/gen uuid?))))
\end{minted}
    \caption{A string uuid spec}
    \label{fig:string-uuid}
\end{figure}

\subsection{Generating specifications of the definitions}
To be able to use the spec references of parameters and responses of complex objects, for input generation and response validation, the actual specs need to be created. 

The specs are created based on the \textit{definitions} part of the OpenAPI document.
The definitions specify the types used by other parts of the document, such as parameters to HTTP verbs or expected responses.
The definitions are specified using the JSON Schema specification \cite{jschema}. Given that the definitions are defined in JSON Schema, implementation is needed to translate JSON Schema to Clojure.spec specifications.

Figure \ref{fig:jschmea-spec} shows an example of how the EDN representation of the definitions part of the document in Figure \ref{fig:open-api-doc} is used to produce specs. Lines 1 - 8 show the call to the \texttt{definitions-to-spec} function with two arguments. The first argument is a string naming a namespace for the resulting specs, used to avoid naming conflicts. The second argument is the EDN-data produced by parsing the JSON Schema.

The result of this function call are three specs: One for each property of the object and one that ties these properties to the object. The specs for the properties, lines 10 - 12 in figure \ref{fig:jschmea-spec}, reference the predefined specs for that type and format. The spec defining the object, starting in line 13, states the name of the spec, \texttt{:definitions/ObjectInfo}, and that the object is a collection of \texttt{keys} followed by the expected name of the keys, which also is a reference to its spec. 

When the specs have been created for all definitions in the document, the pieces needed to be able to generate input data and to validate responses are in place.

\begin{figure}
    \centering
\begin{minted}{clj}
(definitions-to-specs
  "prefix"
  {:ObjectInfo	  
    {:type "object",
     :properties
       {:objectId {:format "uuid", 
                   :type "string"},
	:name     {:type "string"}}}})
=>
((s/def :prefix/objectId 
        :openapi.specs/string-uuid)
 (s/def :prefix/name :openapi.specs/string)
 (s/def 
   :definitions/ObjectInfo 
   (s/keys :req-un [:prefix/objectId 
                    :prefix/name])))
\end{minted}
    \caption{Edn representation to specs}
    \label{fig:jschmea-spec}
\end{figure}

\subsection{Make test generators}
To automatically produce examples of the request parameters in the document, generators are required.
According to the OpenAPI specification, parameters can be of five different kinds. Those are \textit{Path, Query, Header, Body,} and \textit{Form} \cite{openapi-spec-2}. The tool's request generator must take those different kinds into consideration since the HTTP request will look different for different kinds of parameters.

For example, if the type of the parameter is \textit{query}, the generated parameter value should be included in the URL as \textit{/objects?id=value} where \textit{id} is the parameter. In the case of a \textit{body} parameter nothing changes in the URL but the generated parameter value should be included in the \textit{body} value of the HTTP request.

When making the generators we use the specs that have been attached to each parameter. We can leverage that Clojure.spec has the ability of producing TestCheck compatible generators from specs. Figure \ref{fig:generator} shows an example of how straight forward it is to generate examples when we can use the specs created in previous steps.

The URL generators will iterate over each parameter and insert the generated value in the URL (\textit{Path, Query}) or as attached data (\textit{Header, Body, Form}). To increase the likelihood of generating edge-cases the generators can generate data that is outside the specification. In the OpenAPI document, parameters can be marked as \textit{required}. This will be reflected in the specs produced and such values will always be included in the generated value. However, the generators can be configured to not include required parameters. This allows the generators to produce test cases that test the SUTs ability to handle missing input. Further, generators can also be configured to produce values out of range.

The result of this step are URL generators that can randomly produce valid URLs to call the API with random input.

\begin{figure}
    \centering
\begin{minted}{clj}
(gen/sample (s/gen :definitions/ObjectInfo))
=>
({:objectId "2acf532d-c60e-472c-8530-c3b5d138327b", 
  :name ""}
 {:objectId "3cb0fcbe-2f6f-47c6-83de-bccca7c3fdc3", 
  :name "Ax"}
 {:objectId "148aab47-61a8-4132-8d79-2572c015b822", 
  :name "WNDP0"})
\end{minted}
    \caption{Generation of data from spec}
    \label{fig:generator}
\end{figure}

\subsection{Check properties} \label{build-properties}
With all the specifications in place for the documents parameters, responses and definitions, as well as the generators for the URLs, the last step is to build the test properties and check them.
The form of the properties are that \textit{for all} generated URLs for an endpoint, the responses should: 
\begin{itemize}
    \item have a non 500 status code
    \item have a body payload that conforms to the defined spec
    \item have a status code that is defined in the document
\end{itemize}{}

To be able to use previously gained results, stateful properties can be used. Two ways of implementing stateful properties are to either feed the current state to the generators or to use the current state as a guard. 

If the state is given to the generators, the next generated input will be randomly selected from the current state. As an example, a stateful generator could be given a list of valid objects to perform a lookup operation on and return a random item from that list. In the second case a model such as an FSM can be used to verify if the generated data is valid given the current state in the model. If it is not valid it will be thrown away and a new value will be generated.

To be able to generate requests that contain valid input we have used the approach where we give previous results as input to some generators. This is explained further in Section~\ref{being-inteligen}.

When using a property-based approach, testing the SUT is the process of \textit{checking} that our defined properties hold for the generated URLs and input data. The process of one test run is then to map the \textit{check} function on all our URL generators. Given the generated URL the API will then be called. The received result will be checked for the properties defined. The number of tests generated for each end-point is easily configured.

\subsection{Stateful sequences} \label{being-inteligen}

It quickly becomes apparent that to be effective for stateful systems and to be able to have status code coverage our approach needs to be more intelligent than what only randomly generated input will give. This is also a finding of Atlidakis et al. \cite{Atlidakis:2019:RSR:3339505.3339600}.
For example, if looking up an object is done via  a path parameter with the URL template   \textit{/objects/\{id\}} and  the parameter is specified as a \textit{UUID}, then the probability of randomly generating a \textit{UUID} that is identical to an object in the system is very low. Most likely, one would get only \textit{404} response codes (object not found), and never cover the case of looking up an actual object in the system (status code 200).

Our approach is to start out with a pure random input and if we get any result, creating new resources or finding existing ones, store that as a source for future input generation. This means that if we generate a search string that gives a result that was not empty, we store the objects returned. At the next step when generating data for the endpoint that requires an \textit{id} we search the stored results for entities that contain the attribute \textit{id}, if we find any, those entities are then used as a source for our generator. In our implementation we simplified to only support objects with \textit{id} as identifier but the method can be generalized with more engineering effort.

We then start out with pure random input, which is also useful as fuzzy testing, but as we are successful in searches, we learn valid objects that are stored in the system.

\subsection{Report the test results}
To be useful to practitioners the presentation of testing results is important. To save time it should be easy to see the overall result of the test run, but at the same time enough details should be provided to debug any failing test. The default output of TestCheck gives a succinct view of the result. Figure \ref{fig:output} provides an example of a successful and failed result. When a test fails, TestCheck will report the failed input and the smallest example after shrinking.

In addition to the output from TestCheck, our implementation collects data on each test executed. Each API call is recorded, and each returned result is also stored. This information can be used by developers when debugging any failures found. In addition to serving as debug information of all calls made, it is also used to present a frequency table of each endpoint and its return codes. This table shows the URL, the verb called, the response code, and how many calls had this return code. It is thus possible to analyze if all endpoints were tested and if all expected return codes were covered.

\begin{figure}
    \centering
\begin{minted}{clj}
;; Successful result
{:result true,
 :pass? true,
 :num-tests 100,
 :time-elapsed-ms 8607,
 :seed 1558358174968}

;; Failed result
{:shrunk
 {:total-nodes-visited 10,
  :depth 5,
  :pass? false,
  :result false,
  :result-data nil,
  :time-shrinking-ms 791,
  :smallest
  [{:url
   "http://service/api/v1/objects?query=<...>",
   :body []}]},
 :failed-after-ms 5666,
 :num-tests 38,
 :seed 1558358168511,
 :fail
 [{:url
 "http://service/api/v1/objects?query=<...>"
   :body []}],
 :result false,
 :result-data nil,
 :failing-size 37,
 :pass? false}
\end{minted}
    \caption{Default output from TestCheck}
    \label{fig:output}
\end{figure}

\section{Evaluation} \label{evaluation}
To evaluate the proposed approach, a multiple case study was undertaken. The research questions we investigated were:

\begin{itemize}
    \item {RQ1: How do different stateless generators compare in terms of response code coverage and fault finding?}
    \item {RQ2: How do stateful generators compare to stateless generators in terms of response code coverage and fault finding?}
    \item {RQ3: Which additional insights, supplementary to fault finding, can the approach provide?}
\end{itemize}

To answer these questions, we developed a proof-of-concept tool, QuickREST, that implements the proposed approach and applied it in our experiments on several services developed in industry and one open-source project.

\subsection{Studied Cases}
The cases selected to evaluate the proposed approach were a collection of services developed as a back-end for a mobile application in industry, and the open-source development management service \textit{GitLab}. The mobile back-end was selected since it was available at our industry partner and that it was composed of services with a REST API described with OpenAPI. The reason for selecting GitLab as a testing target is to be able to compare our approach with another approach of testing REST-APIs (i.e., using \textit{RESTler} \cite{Atlidakis:2019:RSR:3339505.3339600}).

\subsubsection{Mobile back-end}
The system has two main parts. The client side is a mobile application running on either Android or iOS. The main use cases for the mobile application are to search for process objects, such as motors or pumps, and to display information of these objects. The information ranges from name and type to trends of live run-time values.

The mobile application connects to a back-end that is specifically designed to serve the mobile application. The back-end consists of several services that work together to solve the mobile client use cases. These services are web-services and expose their APIs via REST. The API is also described using OpenAPI. These are the services that we have been using as our system under test.

\subsubsection{GitLab}
As stated by GitLab, "GitLab is a single application for the entire software development lifecycle. From project planning and source code management to CI/CD, monitoring, and security" \cite{gitlababout}. GitLab can be run both as an on-premise solution or as a Software-as-a-service (SaaS) hosted by GitLab. GitLab is available both as a community edition (GitLab CE), which is open-source, and a enterprise edition (GitLab EE), which is closed-source.

GitLab has an extensive REST API \cite{gitlabapi}. However, there is no official OpenAPI document describing the API. For our evaluation, we have manually produced an OpenAPI document including the end-points selected for inclusion in the experiments. The selection was based on the API operations evaluated by RESTler \cite{Atlidakis:2019:RSR:3339505.3339600}. With that selection we knew of the existence of a bug \cite{gitlabsamplebug}, requiring a stateful interaction of first creating a resource, delete the resource and then edit the deleted resource, all within a very short period of time. With the knowledge of this bug we could try to reproduce the findings of RESTler with our method.

We ran GitLab CE locally via a monolithic Docker\footnote{https://www.docker.com/} file \cite{gitlabdocker}. Being monolithic means that the different services that make up a GitLab installation run as one node. This is not a setup for high performance, compared to installing GitLab on a cluster of servers, but it is a setup that is reasonable for a developer to test on as part of their development workflow, thus making it interesting.

\subsection{Experimental Setup}
In the evaluation, we exercised all services of the described mobile back-end system as well as the selected GitLab end-points. Each evaluation run of a service consisted of each individual end-point being tested with randomly generated input. Each end-point was exercised with 10 tests per iteration and after 30 iterations the number of tests were increased by an order of magnitude. The increase in the number of test cases per iteration allows for larger sizes of the generated input. The returned status codes were used as a coverage criterion and included in the test reports. We wanted to cover all status codes defined in the OpenAPI document.

To enable evaluation of RQ1, different frequencies for basic type generators were explored to observe if this changed the number of bugs found. In the first iteration of evaluation, alphanumeric strings were used as the basis for random string generation. The reason being that we would generate valid URLs. In later iterations any string was used to also include invalid input. To further evaluate the effect of different generator frequencies two end-points, where input validation bugs had been found in GitLab, were selected. One of these end-points required string input and one required integer input. 

To evaluate stateful interactions compared to stateless interactions (RQ2) sequences of API calls were generated and executed. The result of each call, if it had a body with a payload, were stored in an in-memory database. This made previous results available for input selection of coming calls in the sequence. For example, if the first call in the sequence was a successful POST operation that returned a new resource, that result was stored. If the second operation was a DELETE that required an id, a search would be performed in the database for an id property. If found, that result was used as input. This made sure that we could evaluate valid stateful sequences. If no previous results were present in the database, input was generated in the same way as for stateless tests. To evaluate if the stateful generators could find faults we aimed to reproduced a known bug in GitLab found by RESTler \cite{gitlabsamplebug}, requiring a stateful interaction.

In the experimental setup each endpoint of the service described in the OpenAPI document was individually tested. Specs were derived from the OpenAPI document and then used as generators. The generator for each endpoint was then given to TestCheck to validate if the defined properties would hold. The properties would check for status codes that do not indicate a crash, that the body of the response was valid according to the specification, and that the returned status code in the response actually is specified in the OpenAPI document. The last two properties were used to evaluate RQ3, where any failed check on those properties indicate a misalignment between the SUT and the OpenAPI document.

\subsection{Results}

\subsubsection{RQ1: How do different stateless generators compare in terms of response code coverage and fault finding?}

\begin{table}
\caption{Generator efficiency for string validation bug}
$T_{fail}$ show the percentage of \textit{iterations} that found the bug using either a generator producing any string or a generator only producing alphanumeric strings. Each setup was iterated 30 times.
\label{table:string-gen}
\centering
\begin{tabular}{c r r}
\hline
\bfseries Test cases / Iteration & \multicolumn{2}{c}{\bfseries Generator $T_{fail}$} \\
& Strings & A/N Strings \\
\hline
100 & 0.0\% & 0.0\% \\ 
1000 & 13.3\% & 0.0\% \\ 
10000 & 63.3\% & 0.0\% \\
\end{tabular}
\end{table}

\begin{table}
\caption{Generator efficiency for int validation bug}
$T_{fail}$ show the percentage of \textit{iterations} that found the bug using either a generator producing any integer or a generator only producing natural integers (\textgreater0). Each setup was iterated 30 times.
\label{table:int-gen}
\centering
\begin{tabular}{ c r r}
\hline
\bfseries Test cases / Iteration & \multicolumn{2}{c}{\bfseries Generator $T_{fail}$} \\
& nat-int & int \\
\hline
10 & 0.0\% & 0.0\% \\
100 & 93.3\% & 60.0\% \\ 
1000 & 100.0\% & 100.0\% \\
\end{tabular}
\end{table}

\begin{table}
\caption{Response code coverage of POST method}
Using the POST method of an API path with a known bug coverage should include 201 (resource created), 400 (bad request) and 500 (internal server error). 100 test cases / iteration for 30 iterations.
\label{table:response-code-coverage}
\center
\begin{tabular}{ c r r r r r}
\hline
\bfseries Response & \multicolumn{5}{c}{\bfseries Generator P(Str, A/N Str)} \\
\bfseries Code & \{1, 0\} & \{0, 1\} & \{0.5, 0.5\} & \{0.3, 0.7\} & \{0.1, 0.9\} \\
\hline
201 & 0.0\% & 86.5\% & 15.7\% & 30.5\% & 63.0\% \\
400 & 83.6\% & 13.5\% & 71.0\% & 55.2\% & 31.8\% \\ 
500 & 16.4\% & 0.0\% & 13.3\% & 14.3\% & 5.2\% \\ 
\end{tabular}
\end{table}

In the experiments, we defined bugs as test cases that resulted in a 500 status code. We found several new bugs during the experiments using a fully automatic approach, both in the mobile back-end and in GitLab \cite{gitlabsamplebug-2, gitlabsamplebug-3, gitlabsamplebug-4, gitlabsamplebug-5}.  The tested APIs had no problem in handling random input of alphanumeric strings. However, when the string generator was changed to include any string\footnote{character values of 0-255}, not only alphanumeric strings, bugs were quickly found.
This set of bugs was produced without any statefulness, i.e. without taking previous interaction results in consideration, or generating sequences of API calls. Since these bugs were not dependent on any sequence of calls or specific states, all bugs were categorized as input validation bugs.

As described in the setup, two known bugs in GitLab were selected to evaluate the ability to find bugs by the different type of generators. Table \ref{table:string-gen} shows the effectiveness in finding the input validation bug where a string was required and Table \ref{table:int-gen} where an integer was required. Looking at the test results of the string and integer generators, it is reasonable to think that we should always use the string generator and the nat-int generator since those were most effective in finding the input validation bugs. However, in addition to finding faults, effective generators should produce input that result in coverage of all defined response codes. Table \ref{table:response-code-coverage} presents the frequencies of different response codes for a POST method in GitLabs API, when tested with generators with different probabilities of strings and alphanumeric-strings. In Table \ref{table:response-code-coverage}, we can see that the \textit{string} generator is useful for finding the input validation bug (500) but will not generate any input that is valid for the end-point, no 201 responses. To achieve full coverage of the defined response codes a mix of generators had to be used.

The automatic generation of generators might then, depending on the implementation, produce generators that do not get full response code coverage. To mitigate this, it is valuable to allow the user to tweak the generators or let the implementation learn a distribution that results in full coverage. In summary, 
\textit{care should be taken when implementing automatic generators to produce the intended range of input values.}

\subsubsection{RQ2: How do stateful generators compare to stateless generators in terms of response code coverage and fault finding?}

We reproduced the known bug in GitLab that required stateful interactions \cite{gitlabsamplebug}. A stateful interaction requires stateful generators, that use data previously seen as a basis to generate new input. Therefore it was not possible to find those kinds of faults with only stateless generators.

Stateful generators greatly increased the likeliness of getting response code coverage on API operations that are dependent on some existing state. Covering a successful response code of a DELETE operation was substantially simplified by first creating the resource with a stateful generator. To get the same coverage with a stateless generator an existing UUID must be generated, and that probability will be low.

While stateful generators simplify covering some cases, they bring their own set of challenges. When random sequences of API calls are generated, random input can be used. But to be able to perform multiple operations on the same resource, the identity of the resource should be selected from a known source, as a database or model. To be effective, stateful properties can use random input for the creation of new resources and random sequences of API calls, but guided input to be able to perform multiple operations on the same resource.

In the cases where the identity of a resource is not consistent in an OpenAPI document, a mapping might be required from the user. For example, if a POST request on \textit{api/persons} is performed and the result is a JSON object with the identity property of \texttt{personId} but the DELETE end-point is specified as \textit{api/person/\{id\}}, we need some input to make that connection. Hence, stateful generators are limited to how well the identity relationships are expressed in the specification. This is also an area where the implementation could apply learning to realize implicit resource relationships in complex systems.
In summary, 
\textit{random input is effective for both stateless and stateful properties but need to be guided for stateful resource selection.}

\subsubsection{RQ3: Which additional insights, supplementary to fault finding, can the approach provide?}
We found that several of the service APIs were under-specified. Consequently, the tests produced a failure when a response code is received that could not be found in the given OpenAPI document. The lack of specification was both for \textit{responses} and \textit{parameters}. For the responses the tool could automatically detect this since it is expressed in the properties checked during testing. To find the under-specified \textit{parameters} we had to check the log files of the SUT. An example of an under-specified parameter was one that was specified as an array of \textit{strings} but, to be valid input to the SUT, it had to be an array of \textit{strings} with the format of \textit{UUID}.
From the tools perspective, it got a 400 status code (i.e., malformed input) but the tool could not infer that this is due to lack of specification rather than random input. Looking at the logs of the SUT made the problem apparent, since the input was logged as not conforming to UUIDs.

The UI of the SUT was used after running several tests. This revealed that in some places very long strings, that had been produced as random input, were not truncated and made the UI look bad.

Most of these problems were found by observation by a human. It could be argued that this then could be found without any tool. But we found that the tool is an \textit{augmentation} to a human doing exploratory testing. The human tester does not have to manually produce test cases but can run the tool and act as an observer. 

\section{Related work}
Property-based testing has been used to find faults with success on real industry systems, in multiple domains, such as telecom systems \cite{Arts:2006:TTS:1159789.1159792}, file synchronization services \cite{7515466}, automotive systems \cite{7107466} and databases \cite{Hughes2016}. 

For web services, as we target, PBT have been used for services implementing Web Services Description Language (WSDL) specifications, translating the specification to generators and properties \cite{Francisco:2013:TWS:2505305.2505306, PropErWebServiceTesting, Li:2014:APT:2543728.2543741}. For REST services Seijas et al. proposed PBT, however according to the authors this was a highly manual approach \cite{LamelaSeijas:2013:TPT:2505305.2505317}.

\textit{Jsongen}, is a library for Quviq QuickCheck introduced by Fredlund et al. \cite{Fredlund:2014:PTJ:2680842.2681200} to generate input from JSON Schema to test web services. It was extended by Earle et al. \cite{BenacEarle:2014:JQB:2633448.2633454} to include a service's behaviour, to automatically explore a web service. JSON Schema is related to OpenAPI since parts of an OpenAPI document is expressed in JSON Schema. While JSON Schema is considering only data, OpenAPI also includes the service model.

Aichernig et al. proposes an approach to use business rules in the form of XML, and from those create extended finite state machines used in PBT tests \cite{Aichernig2019}. This is an automatic approach and has some level of intelligence, by walking the FSM, but requires that the services behaviour is available as such an XML artifact.

OpenAPI/Swagger documents have been used to generate tests, both in black- and white-box testing. EvoMaster is a white-box approach that generate tests based on a given Swagger document \cite{Arcuri:2019:RAA:3292526.3293455}. Usage of EvoMaster require some developer implementation effort. To require as little developer interaction as possible and to be agnostic of the target platform and languages used, we have used a black-box approach. 

RESTler introduces a similar approach to ours by using an automatic black-box approach to intelligently fuzzy-test REST APIs \cite{Atlidakis:2019:RSR:3339505.3339600}. As stated by the authors, RESTler aims to be a security testing tool. However our goal is to not only find faults but to explore the given REST API and in doing this we test the services with a wider scope of properties. Both RESTler and our approach use the status codes from the REST API calls as an oracle, but in addition to that we also validate that any payload received actual conforms to the specification in the OpenAPI doc, resulting in a stronger oracle. Our approaches also differ in that we use randomly generated input, not a predefined dictionary.

Ed-douibi et al. propose a model-based approach testing REST APIs specified with OpenAPI \cite{8536162}. This approach is related to ours with the main difference that we use a property-based method for test generation. Our approach does not only generate static test cases but it also randomly generates parameter values and sequences of operations that leverages previously returned results to perform stateful operations. 

\section{Discussion and Future Work}

QuickREST turns out to be a tool that allows developers and testers to easily test and explore REST APIs. Developers will get quick and actionable feedback while developing, with the option to manually tweak the tests. Testers can both use the tool to find faults and also to exercise a system while doing other tests. However, to get more conclusive results QuickREST would need to be applied to a larger set of systems under test. Here we discuss some implications that our results might have for practitioners and for future research. 

\subsection{Developer friendly}
As individual end-point tests can be generated in seconds, it is a very developer friendly approach since it can be used during development without interrupting a continuous workflow. However, as the number of end-points and parameters increase, so does the time required. Hence, larger tests may be more appropriate for a continuous integration server.

\subsection{Shrinking}\label{sec_shrinking}As described in Section \ref{property-based testing}, \textit{shrinking} of generated random input is a common feature of PBT. We experienced this firsthand during our experiments. Here we describe two examples of shrinking of the input that produced two of the bugs described. One \textit{path} parameter and one \textit{body} parameter.

Figure \ref{fig:pre-shrinked-input} shows a piece of an example of pre-shrinked generated input as a \textit{body} to be used in a \textit{POST} request. The result of the automatic shrinking process is shown in Figure \ref{fig:post-shrinked-input}. Just by comparing the generated input from before and after shrinking it is apparent that the smaller input is preferable as a reproducing case for developers.

In the case of a failing \textit{Path} example, parameters contains unprintable characters so we do not include a verbatim example. The form of an url with path parameters is \texttt{/api/v1/objects?name=<random input here>}. In our case the shrinking process could shrink the random parameter from 20 characters down to 2 characters.

The shrinking feature of PBT is very useful for developers who want a reproducing case as small as possible. However, for a fully automatic approach you can get "stuck" with shrinking. For example, if we have an end-point with input bugs in several of the parameters, shrinking will bring that down to the first case. This means that to get further in our testing we have to exclude that parameter. This might require manual effort, thus breaking the full automation.
However, it could be argued that this kind of shrinking problem forces an agile approach where a developer needs to fix the first found bug before continuing with further testing.

\subsection{Mutate the specification}
In an OpenAPI document, parameters can be specified as \textit{required}. To produce valid input to the API this needs to be respected. But it is useful for an automatic approach to in some cases leave out required parameters or change the specified type, \textit{mutating} the API specification. This will produce test cases for a proper input validation of not only the value of a parameter but if the parameter itself exists or not.

\subsection{Iterate over parameters}
The size of the input domain covered will be dependant of how many parameters are included. Given an end-point with, for example, 5 required parameters, and 100 tests. Each test case will generate input for all parameters and thus after 100 iterations we have not gone as deep in each parameter as would be the case if there were only a single parameter. Therefore, an automatic approach would get larger input coverage by producing tests for each parameter before starting to use combinations of parameters.

\subsection{Before going stateful, make sure API handles input}
Generating stateful interactions will not be effective in finding stateful bugs if the SUT does not handle generated input. If the SUT does not handle input validation any sequence of interactions will shrink to just one call, the one with bad input. We recommend starting with stateless input generation and when the API can handle that, add stateful sequences.

\subsection{Stateful shrinking might not be accurate}
When shrinking is performed the API will be called, changing the state of the system. This can result in that the shrunken output is not actually exactly what will produce the error. However, we observed that in our cases, the \textit{sequence} was correct but not the \textit{input parameter values}. In that case, consulting the logs from the system will show the actual input used in the sequence.

\subsection{Augment with manually created model}
An automatic approach is a good way to get a lot of testing done with a low effort. However, to get maximum leverage out of this kind of approach a model could be helpful in guiding stateful interactions. Such a model then keeps track of the expected state of the system, guiding valid operations and is a source of verification of how the system should look like.

\begin{figure}
    \begin{minted}[frame=single,linenos, xleftmargin=8pt, framesep=1mm, fontsize=\scriptsize, numbersep=2pt]{clj}
{:body
[ ...
  ...
 {:variables #:c0{:T \F},
  :description "x",
  :objectId "077de3d9-3f50-4756-bb45-ee61be31e8a2",
  :model "",
  :type "",
  :name "v7"}
 {:properties {},
  :variables {\@ "\""},
  :description "8",
  :objectId "d9bd22ce-939c-4332-8702-0430e38f04c5",
  :model "s0",
  :type "",
  :name "0O"}
  ...
  ...]}
    \end{minted}
    \caption{Abbreviated pre-shrinked input (60 lines)}
    \label{fig:pre-shrinked-input}
\end{figure}

\begin{figure}
    \centering
    \begin{minted}[frame=single,linenos, xleftmargin=8pt, framesep=1mm, fontsize=\scriptsize, numbersep=2pt]{clj}
{:body
[{:variables {1.0 0},
  :objectId "ec9c007f-5278-45a7-8aa2-b4cb13ce9e11",
  :model "",
  :type "",
  :name ""}]}
    \end{minted}
    \caption{Complete shrinked input}
    \label{fig:post-shrinked-input}
\end{figure}

\section{Conclusion}
We have introduced a method and a proof-of-concept implementation to automatically test REST-APIs described with OpenAPI by leveraging PBT. The described approach leverages existing libraries and techniques and to the best of our knowledge this is the first approach that uses automatic PBT in combination with OpenAPI documents to test REST APIs, with the intent of not only finding faults but also to learn more about the SUT. The experimental results on a real industry software system show that this approach can find real faults and help in gaining new knowledge of the system with a very low effort from the developers and testers.

PBT is a useful technique with substantial tool support that can be leveraged in industry. For example, the capabilities of Clojure's dynamic data processing in combination with Clojure.spec and TestCheck was shown to be a powerful tool-chain to automate our method. The act of shrinking is indicated as a rather useful feature for industry. 

In our experience, the usage of PBT in industry seems rather limited. This is unfortunate since, as we have shown, it is an approachable technique that can find real bugs and can be extensively utilised in part of a developer workflow. 

To make this approach more effective and more suitable for deeper exploration, we could augment it by a model of the call sequences a real user would perform. In addition, it would be useful to help humans by automatically analyzing the logs while running tests. Exploratory testing is an important part of ensuring the quality of software systems but automation in this area is lacking although there are opportunities for it. The results presented in this paper show that it is possible to assist humans in exploration of REST-APIs with QuickREST.

\section*{Acknowledgements}
This work is supported by ABB, the industrial postgraduate school Automation Region Research Academy (ARRAY) funded by The Knowledge Foundation. Additional support is provided by ITEA3 TESTOMAT project funded by VINNOVA.

%\newpage
\bibliographystyle{IEEEtran}
\bibliography{main}
\end{document}